# From Metadata to Storytelling: A Framework For 3D Cultural Heritage Visualization on RDF Data

Sebastian Barzaghi[1], Simona Colitti[2], Arianna Moretti[3], Giulia Renda[4]

[1] Department of Cultural Heritage, University of Bologna, Italy - sebastian.barzaghi2@unibo.it
[2] Department of Architecture, University of Bologna, Italy – simona.colitti2@unibo.it
[3] Department of Classical Philology and Italian Studies, University of Bologna, Italy – arianna.moretti4@unibo.it
[4] Department of Classical Philology and Italian Studies, University of Bologna, Italy – giulia.renda3@unibo.it

**ABSTRACT (ENGLISH)**
This paper introduces a pipeline for integrating semantic metadata, 3D models, and storytelling, enhancing cultural heritage digitization. Using the Aldrovandi Digital Twin case study, it outlines a reusable workflow combining RDF-driven narratives and data visualization for creating interactive experiences to facilitate access to cultural heritage.

**ABSTRACT (ITALIANO)**
*Dai Metadati allo Storytelling: Un Framework per la Visualizzazione del Patrimonio Culturale 3D su Dati RDF.* Questo articolo presenta una pipeline per integrare metadati semantici, modelli 3D e narrazione, migliorando la digitalizzazione del patrimonio culturale. Utilizzando il caso di studio del gemello digitale di Aldrovandi, introduciamo un flusso di lavoro riutilizzabile per combinare narrazioni e visualizzazioni su dati RDF per rendere il patrimonio culturale più accessibile attraverso esperienze interattive.
**Parole chiave:** Gemelli Digitali; Storytelling; Esperienza dell'Utente; Visualizzazione Dati; RDF

## 1. INTRODUCTION
The CHANGES project[1] ("Cultural Heritage Active Innovation For Next-Gen Sustainable Society") is an EU-funded initiative to promote, preserve, and make accessible Italian cultural heritage (CH). Its Spoke 4 focuses on using virtual technologies to enhance CH in museums and art collections. Engaging cultural institutions in this effort improves artefacts' knowledge, curation, and management, while also expanding public involvement through innovative communication strategies (Balzani et al., 2024). While RDF is a standard for managing cultural heritage metadata, significant gaps persist in its application to 3D objects, particularly in leveraging metadata to enhance user experiences and engagement. The limitation is critical in museums, where integrating metadata with narrative techniques has shown great potential to enrich visitors' engagement (Grasso et al., 2024). Both physical and digital storytelling approaches combined with data visualization have proven effective in conveying knowledge and fostering interaction with CH collections (Fernie et al., 2012). Nonetheless, no comprehensive framework already bridges the gap between RDF metadata and 3D objects through narrative techniques. This paper explores this gap by reviewing relevant works (Section 2), proposes an interdisciplinary and reusable workflow to address it (Section 3), and demonstrates its application in the Aldrovandi case study (Section 4).

## 2. STATE OF THE ART
### 2.1 SEMANTIC METADATA FOR 3D OBJECTS
In CH digitization, capturing metadata about the cultural heritage object (CHO) and the process is required for management, enrichment, and presentation in a virtual environment. In recent years, several projects have explored semantic approaches for 3D CH metadata. Early efforts, like 3D-ICONS (D'Andrea & Fernie, 2013), focused on integrating 3D models into Europeana using CIDOC-CRM and CRMdig extensions. In the following years, widespread ontologies were largely adopted for modeling RDF metadata of CH objects, in projects such as the one by Messaoudi et al. (2018), who created a pipeline to integrate various information about built heritage into an ontological model aligned with CIDOC-CRM and other related modules. Catalano et al. (2020) later developed a methodology to include quantitative documentation data, expressed according to a schema based on CIDOC-CRM and CRMdig, in the description of 3D CH artifacts. Recently, the trend of reusing and extending existing models to describe 3D CH data re-emerged with a focus on reproducibility. Amato et al. (2023) illustrated the metadata management process used to select elements from existing models and optimize their reuse, while Hermon et al. (2024) proposed a framework to integrate a CHO's data from several sources into a digital twin, including its data, documentation, and digital model.

---
[1] https://www.fondazionechanges.org/ (cons. 24/01/25)

To make real use of metadata across individual projects, a specific effort should be devoted to ensuring their FAIRness, focusing on making data interoperable. To this end, an extensive crosswalk mapping (European Commission. Directorate General for Research and Innovation. & EOSC Executive Board, 2021) was proposed to facilitate data exchanges. A key practice for interoperability is publishing datasets as Linked Open Data (LOD), using code-oriented tools for semantically structuring data, such as the Python library RDFLib, although inaccessible to users who can't code. Alternative approaches are based on mapping tools, such as the RDF Mapping Language (RML) (Dimou et al. 2014), a Java-based engine limiting the interaction to configuration files only, without the necessity of programming, unless specific conversion necessities emerge. Based on RML, Morph-KGC (Arenas-Guerrero, Chaves-Fraga, et al., 2024) was developed to allow Python programmers to extend RML functionalities with user-defined functions (Arenas-Guerrero, Espinoza-Arias, et al., 2024).

**2.2 STORYTELLING, DATA VIZ DESIGN, USER EXPERIENCE DESIGN**
Design-driven approaches in the digitization of CH allow for transforming collected data into accessible and meaningful resources by integrating contributions from both design and humanities disciplines (Lupo et al., 2023; Celi et al., 2016). Storytelling through data-informed approaches offers benefits in interpreting and engaging with CH collections, while information visualization enhances access and understanding with representations that simplify complex datasets. In addition to that, integrating RDF metadata allows for exploiting semantic connections in the narratives. This methodology not only increases accessibility in compliance with inclusive standards but also fosters data reuse through open-access licenses, benefiting both general audiences and experts. A systemic design approach ensures swift access, effective technology transfer, and cultural relevance. Combining innovative technologies with open access principles, the design fosters broad, sustainable, and interdisciplinary engagement, enriching CH through community and stakeholder interaction with advanced digital tools. Designers play a crucial role in facilitating communication among heritage experts, technologists, and end users by translating diverse needs and expectations into practical, user-friendly solutions that enhance the accessibility and meaningfulness of cultural heritage (Trocchianesi & Bollini, 2023). The process relies on advanced metadata modeling, prototyping, and accessible outputs such as data stories. These narratives make complex datasets comprehensible, engaging, and reusable for diverse audiences. Data storytelling has been shown to bridge the gap between data complexity and user experience, enabling new forms of engagement and contributing to collective digital memory (Masud et al., 2010). By leveraging tools like RawGraph[2] and Figma[3], pre-prototyping phases facilitate the exploration of visualization strategies that reinforce the project's visual identity (Mauri & Ciuccarelli, 2012).

**2.3 SUCCESSFUL STORYTELLING STRATEGIES ADOPTED FOR OTHER DIGITAL RESOURCES**
The exploration of data-driven design methodologies for CH has been the key focus of the Polifonia project[4], a European H2020 initiative dedicated to preserving and disseminating music heritage through innovative digital tools. As part of its contributions to the field, Polifonia developed a novel methodology that combines eXtreme Design (Carriero et al., 2021; Presutti et al., 2009) and Design Thinking (Chasanidou et al., 2015; Rowe, 1991) to bridge the gap between ontology-driven approaches and user-centered interface design (Grasso et al., 2024). This hybrid framework ensures that domain-specific requirements, as well as user needs, are seamlessly integrated into the development of tools for information retrieval, exploration, and dissemination (Renda, Grasso, et al., 2023). By integrating these structured insights into design workflows, Polifonia has established a replicable model for creating applications that cater to diverse audiences while leveraging the rich, structured data offered by LOD. Importantly, the tools developed through this methodology have undergone evaluation and are already being used in real-world scenarios. This highlights the practical reliability and effectiveness of the approach, reinforcing the potential for reuse and adaptation.

---

[2] https://www.rawgraphs.io/ (cons. 24/01/25)

[3] https://www.figma.com/ (cons. 23/01/25)

[4] https://polifonia-project.eu/ (cons. 26/01/25)

One of the key products of this methodology is MELODY[5], an open-source platform that empowers users to author content in the form of *data stories* (Renda, Daquino, et al., 2023). Designed to promote accessibility and engagement with LOD, MELODY enables domain experts and general users to combine data-driven visualizations with curated narratives. It provides a user-friendly *What You See Is What You Get* (WYSIWYG) interface for creating, editing, and publishing articles that integrate SPARQL query results as visual components such as charts, maps, and tables. MELODY's narrative-driven approach addresses significant challenges in presenting LOD. The tool supports the creation of content that combines interpretive storytelling with data visualization, making complex datasets accessible and engaging to a broader audience. Its flexible, open-source architecture ensures that it can be reused with any SPARQL endpoint. Additionally, its open-source nature allows users to extend or modify the platform according to their specific needs, making it an adaptable solution for a wide range of applications in the CH domain.

## 3. METHODOLOGY

Our research adopts a user experience-oriented, iterative, and multidisciplinary approach to tackle the challenge of enhancing access to CH resources through digital means. This approach is characterized by multidisciplinary collaboration among researchers from diverse fields, ensuring a holistic solution. Open data principles guide the entire process, making all outputs reusable and scalable within the LOD paradigm. A key focus is on narrative techniques, which enhance user engagement through interactive paths within the digital environment. Systematic metadata integration bridges gaps in 3D heritage model accessibility, developing a comprehensive pipeline from data collection to visualization. As shown in Figure 1, this process consists of parallel, iterative activities.

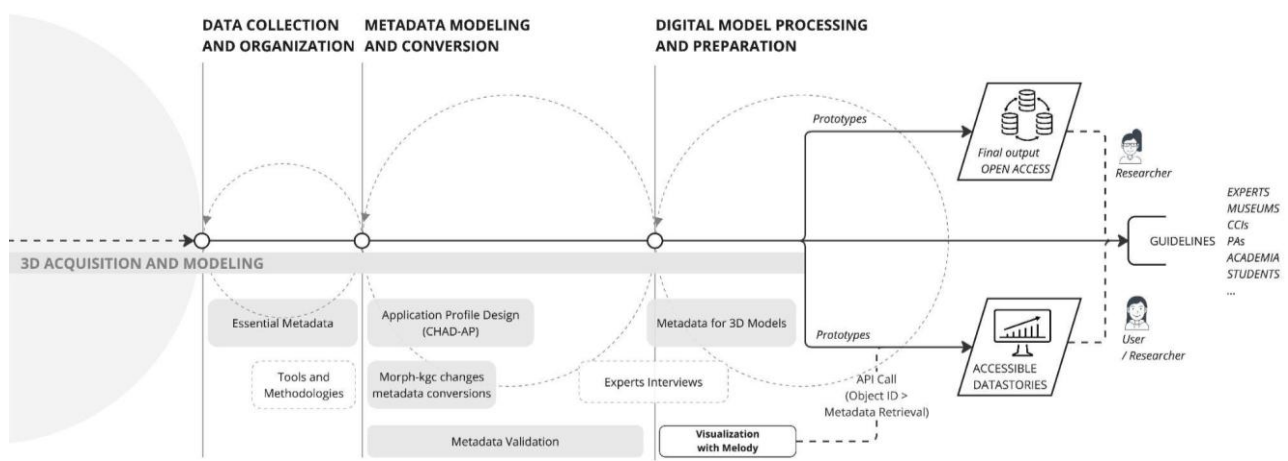

**Figure 1. Iterative process for CH digital accessibility. The figure showcases the methodological framework encompassing iterative UX design, metadata integration, and storytelling techniques aimed at improving the accessibility and engagement of CH resources in digital environments.**

In the **data collection and organization** phase, metadata about CHOs and digitization steps were organized into two Google Sheets[6] tables, chosen for their intuitive interface, real-time collaboration, and versioning. Each column represents a metadata field, and each row corresponds to a unique CHO.
In the **metadata modeling and conversion** step, the spreadsheet-collected tabular metadata was transformed into RDF triples through a two-step process: 1) using an ontology development methodology such as SAMOD (Peroni, 2017) to create an application profile of CIDOC-CRM for interoperability, and 2) converting tabular data into RDF triples using an ad-hoc extended version of Morph-KGC[7], which leverages RML for semantic conversion.

---

[5] https://projects.dharc.unibo.it/melody/ (cons. 26/01/25)

[6] https://workspace.google.com/products/sheets/ (cons. 26/01/25)

[7] https://github.com/dharc-org/morph-kgc-changes-metadata (cons. 10/04/2026)

The **prototyping** phase employs iterative user experience and service design principles. Using Figma, prototypes were refined through stakeholder feedback, combining collection-level narratives with object-specific stories to explore connections between CHOs. This approach enhances accessibility, engagement, and adaptability through intuitive interfaces linking metadata insights with storytelling.

For **digital model processing and preparation**, MELODY was used to integrate LOD, data visualization, and narrative techniques to engage users. While robust, MELODY's reliance on single SPARQL endpoints and manual processes limits scalability for large-scale reuse. To address these issues and integrate seamlessly with ATON (Fanini et al., 2021)—the 3D visualization framework used in Spoke 4—we developed an API that reuses MELODY's core components to dynamically generate customized data stories for 3D objects, improving scalability and interoperability.

## 4. USE CASE

In this section, we present the Aldrovandi temporary exhibition Digital Twin (Balzani et al., 2024) as an example to illustrate the application of our approach. In 2022-2023, the exhibition "*The Other Renaissance - Ulisse Aldrovandi and the Wonders of the World*" was held at Palazzo Poggi in Bologna, gathering a varied selection of objects from different sources. The exhibition, which is no longer accessible today, has been entirely digitized, including reproductions of interiors and 3D models of each object with accurate metadata, both on objects and the entire digitization process. As a result, the Aldrovandi Digital Twin is a 3D representation of the exhibition, showcasing how our methodology integrates RDF metadata-driven storytelling with virtual technologies to enrich user interaction with CH.

### 4.1 DATA COLLECTION, MODELING AND CONVERSION

The process of digitizing the exhibition's CHOs (Barzaghi, Bordignon, Collina, et al., 2024) started by capturing 3D data through structured-light scanning and photogrammetry, according to the object's physical characteristics and accessibility. Then, the raw data (RAW) was processed and refined into detailed 3D models. More specifically, three versions of each 3D model were retained for practical and scientific purposes: the raw processed model (RAWp), the digital CH object (DCHO), and the optimized version for real-time use (DCHOo). Next, after the models were optimized for performance, they were exported in appropriate formats for real-time web visualization.

In the meantime, two tables were created on Google Sheets: a table was used to collect the descriptive metadata of the exhibition objects, to capture their historical and contextual information, while the other one was used to collect the data of the whole digitization process, recording contextual information for each of the aforementioned phases. While the models were finalized, their metadata in tabular form were exported in CSV format and converted into RDF triples for structured representation. The semantic model used to describe the objects and processes involved in digitizing CHOs in a structured way is an application profile aligned with the CIDOC-CRM framework and called *Cultural Heritage Acquisition and Digitisation - Application Profile* (CHAD-AP)[8] (Barzaghi, Heibi et al., 2024). CHAD-AP has been implemented as an OWL ontology organized into two main modules: the *Object Module* (OM), dedicated to describing the CHO's characteristics and context, and the *Process Module* (PM), for describing the process of acquiring and digitizing it. The conversion of the data contained in both OM and PM into a RDF graph structured according to CHAD-AP specifications was realized by using a custom extension of the Morph-KGC software, which included mapping files, configuration files, and user-defined functions for handling the complex semantic constructs hidden in the tabular format, together with a Python launch script that orchestrated data preprocessing, mapping, and post-processing tasks.

Finally, the ATON framework was used to manage and present the models interactively on the Web, while a SPARQL endpoint was made available for querying and providing structured data about the original CHOs, their digital models, and the whole process of their acquisition and digitization.

### 4.2 PROTOTYPING

As part of an ongoing design research effort, user needs were continuously gathered and analyzed through questionnaires and expert consultations conducted alongside the development process. This iterative engagement identified specific requirements and highlighted the narrative potential of the collections' data, ensuring that the data stories aligned with user expectations and heritage valorization goals. Feedback from stakeholders shaped key elements for structuring meaningful data stories and refining the metadata framework, fostering a dynamic exchange of insights. This process ensured that the data-driven storytelling approach remained relevant and adaptable to the evolving project needs.

---
[8] https://w3id.org/dharc/ontology/chad-ap (cons. 24/01/25)

The creation of data-informed storytelling for CH collections requires a methodological framework that incorporates various visualization techniques and design strategies. This involves several key phases. (1) Metadata modelling and conversion ensure metadata interoperability and quality through structuring, validation techniques, and expert interviews; (2) the selection of visualization and navigation design defines appropriate visualization types (e.g., bar charts, maps, interactive diagrams) and supports the creation of intuitive, immersive interfaces using tools like Figma, with content categorized by themes and visualizations to enhance user guidance; (3) accessibility and usability testing involves rigorous compliance verification using tools like WebAIM[9] for contrast testing and integrating assistive technologies for users with disabilities; (4) standardization and usage guidelines are developed to provide adaptable heritage digitization protocols, including a white paper detailing workflows and data integration methods; (5) monitoring and feedback mechanisms are established to refine data stories and improve project outcomes; (6) impact assessment evaluates accessibility and audience engagement, focusing on integrating digitized heritage into national repositories and expanding the model to additional collections. This iterative, participatory approach ensures that CH is not only digitized but also actively explored and enriched. Visualization tools such as bar charts and interactive diagrams provide an adaptive storytelling experience, facilitating user exploration and navigation through complex heritage data. These tools add value beyond physical exhibitions, fostering deeper engagement and uncovering new connections between collection pieces.

### 4.3 PROCESSING AND PREPARATION

Within the ATON framework, each DCHOo is visualized as a 3D digital twin. When a user selects an object, ATON sends an HTTP request to the API. This request includes the unique global persistent identifier for the Aldrovandi artifact and the location of a configuration file. Mirroring design research and decisions, the JSON configuration file specifies the details to be displayed in the data story, including the object typology, SPARQL queries for metadata retrieval, textual descriptions, visual elements, and interactive components. The API processes the configuration file dynamically executing the SPARQL queries to retrieve metadata from the specified endpoint. For the Aldrovandi artifact, these queries extract information such as its type, materials, dimensions, historical context, place of conservation, and details about its digitization process. This ensures that all relevant metadata is retrieved and contextualized for the data story. Using the metadata retrieved via SPARQL queries, the API generates a customized HTML page that serves as the data story for the 3D object the user is interacting with. This page is designed to include dynamic, artifact-specific sections such as descriptions of the object's materials, historical period, and the conservation site, along with visualizations like charts or images. The design ensures that the data story complements the 3D representation of the object by providing an additional layer of information, allowing users to explore the artifact in depth while engaging with its historical and cultural context.

### 5. CONCLUSIONS

In this contribution, we presented an iterative, reusable, and cross-disciplinary workflow for integrating RDF-driven data stories with 3D virtual representations of CHOs. Following a review of the state of the art, we outlined a methodology that includes data collection and organization, metadata modeling and conversion, prototyping, and model processing and preparation. The workflow was applied to the Aldrovandi Digital Twin use case to illustrate how semantically rich metadata can be used to enhance access and usability of virtual representations of CHOs. By integrating storytelling techniques with visualization strategies, the approach seeks to transform raw data into actionable knowledge while preserving the integrity and authenticity of the heritage materials. The effectiveness of this workflow in producing engaging data stories relies heavily on the quality and richness of the underlying data. Insufficient metadata or lack of contextual detail may limit the ability to craft a meaningful narrative or identify compelling connections between objects. The manual metadata preparation and narrative design aspects may also hinder efficiency and reproducibility in some contexts. Moreover, this work has not yet undergone extensive evaluation as an exploratory effort.

Further developments will also address user experience across different audience profiles. Researchers and domain experts are expected to prioritize access to structured metadata, semantic interoperability, and opportunities for data reuse. In contrast, general users may benefit more from accessible narrative pathways and intuitive, exploratory interfaces. To support both audiences, the integration of panoramic views and synthetic data representations is intended to enable stepwise exploration, from general overviews to detailed insights, aligning with design principles that promote cognitive scalability and content adaptivity. These design choices aim to ensure that the interface not only conveys the richness of the underlying RDF data but also remains accessible and engaging for diverse user groups.

---

[9] https://webaim.org/ (cons. 24/01/25)

Despite the experimental stage of the process, having been developed within the framework of the Project CHANGES, the introduced workflow is meant to be readopted systematically in other case studies, and eventually become the standard for visualizing RDF data in this context. In the short to medium term, this will provide the opportunity to collect and analyze concrete data to evaluate the workflow's efficacy with different input material rather than the dataset it was developed for and tested on until now. The collected feedback will be exploited to iteratively improve the pipeline, to deliver a research tool as useful and easily reusable as possible.  Consequently, future efforts will focus on refining the workflow and testing implementation for the Aldrovandi use case with advanced navigation and search capabilities, prioritizing accessibility and usability testing to ensure an inclusive user experience. Additionally, we aim to evaluate the project's impact on future studies, involving additional case studies (such as other collections and museums) and even situations characterized by sparse or low-quality data. Other interesting venues for further experimentation include testing automatic or semi-automatic approaches to facilitate the data modelling phase and developing quantitative methods to be integrated in the evaluation process that bring together (meta)data quality, service and narrative design heuristics, and the assessment of technical aspects such as data conversion and API processes. Finally, we also plan to work on actionable guidelines that standardize our approach and facilitate its adoption across other cultural heritage projects and institutions.

**ACKNOWLEDGEMENTS**

This work was partially funded by Project PE 0000020 CHANGES - CUP B53C22003780006, CUP J33C22002850006, NRP Mission 4 Component 2 Investment 1.3, Funded by the European Union - NextGenerationEU.